\documentclass[a4paper]{article}

\usepackage{a4wide}
\usepackage{amsmath}
\usepackage{graphicx}
\usepackage{latexsym}

\DeclareMathOperator{\tr}{Tr}

\newcommand{\ie}{\textit{i.e.}}
\newcommand{\etal}{\textit{et~al.\ }}
\newcommand{\eg}{\textit{e.g.\ }}

\title{Gauge-invariance in hadronic scattering processes\footnote{Contributed talk presented at the {\sc{i3hp}} Topical Workshop at the University of St Andrews, Scotland, August~30 - September~1, 2004.}}
\date{}
\author{C J Bomhof\footnote{\textit{E-mail address}: cbomhof@nat.vu.nl}\\[3mm]
{\small\textit{Department of Physics and Astronomy, 
Vrije Universiteit Amsterdam,}}\\
{\small\textit{De Boelelaan 1081, NL-1081 HV Amsterdam, The Netherlands}}
}

\begin{document}

\maketitle

\vspace{-8mm}

\begin{center}
\parbox{0.9\textwidth}{\begin{small}
High energy hadronic scattering processes can be described in terms of 
partonic scattering processes and parton distribution and fragmentation 
functions. These are bilocal matrix elements of quark operators. Colour gauge-invariance requires inclusion of a gauge link, which is characterized 
by its path. For transverse momentum dependent distribution and fragmentation functions, appearing in single spin asymmetries, different paths appear
depending on the process. The difference is related to a time-reversal odd
gluonic pole matrix element. We discuss briefly the path structure of the gauge link in more complicated processes.\end{small}}
\end{center}

\vspace{4mm}

The description of hadronic scattering processes is complicated by several aspects.
Hadrons are not elementary particles, 
but rather bound states of quarks and gluons.
The structure of hadrons in terms of the partons and the processes that bind them together are, 
up to today, mostly unknown.
Obviously, these are all inherently non-perturbative properties of {\sc{qcd}} and cannot be calculated using the method of Feynman diagrams. 
A very intuitive first approach to the treatment of hadronic scattering processes is provided by Feynman's parton model
(Feynman 1969).
The basic assumption of the parton model is that,
on the time scale of the hard scattering process,
we can regard the partons as essentially free particles.
This picture has proven to be a very good first approximation in deep inelastic scattering ({\sc{dis}}: $\ell+h\to\ell^\prime+X$),
the simplest of all hadronic scattering processes.
Schematically, the scattering cross section of {\sc{dis}} in the parton model can be written as
$\mathrm d\sigma^{h \ell}
\propto\sum_q\int\mathrm dx\ f^q(x)\,\mathrm d\hat\sigma^{q \ell}(x)$.
Here $\mathrm d\hat\sigma^{q \ell}$ is the partonic scattering cross section that can be calculated perturbatively and the $f^q(x)$ are the distribution functions of parton species $q$ with momentum fraction $x$.
These distribution functions contain all the non-perturbative information of the hadrons.
Hence they cannot (at present) be calculated from first principles.
However, they can be measured in certain types of experiments and, then, 
be used to make predictions about other hadronic scattering processes.

At leading twist there are three different quark-distribution functions:
the \emph{unpolarised distribution} 
$f_1^q(x)=q(x)$
which measures the distribution of unpolarised quarks of flavour $q$ in an unpolarised hadron,
the \emph{longitudinal polarisation distribution} 
or \emph{helicity distribution}
$g_1^q(x)=\Delta q(x)$
which measures the distribution of longitudinally polarised quarks in a longitudinally polarised hadron,
and the \emph{transverse polarisation distribution} 
or \emph{transversity distribution}
$h_1^q(x)=\delta q(x)$
which measures the distribution of transversely polarised quarks in a transversely polarised hadron.
The functions $f_1^q$ and $g_1^q$ can be measured relatively easy in {\sc{dis}}.
Measurements of $g_1^q$ revealed that the quark helicities only account for about 20\% of the helicity of the hadron.
This `spin crisis' sparked renewed interest in the role of quark-spin in hadrons and the transversity function $h_1^q$.
Transversity, though, 
has proven to be much more difficult to measure than the other two functions. Consequently our knowledge of this function has fallen behind a little bit with our knowledge of the unpolarised and longitudinally polarised distribution functions.
A more complete discussion on these issues can be found in (Ryckbosch 2004).

For a theoretical description of the distribution functions we need a proper definition of these objects.
An appropriate framework is provided by the quark-correlator (Soper 1977)
$\Phi_{ij}(k)
=(2\pi)^{-4}\int\mathrm d^4\xi\:\:e^{ik\cdot\xi}\;
\langle P,S|\,\overline\psi_j(0)\,\mathcal U(0,\xi)\,\psi_i(\xi)\,|P,S\rangle$. 
All the distribution functions are projections of the quark-correlator:
$f_1(x){=}\int \mathrm dk^-\mathrm d^2\boldsymbol k_T
\tr\big[\gamma^+\Phi(k)\big]$, 
$S_L\,g_1(x){=}\int \mathrm dk^-\mathrm d^2\boldsymbol k_T
\tr\big[\gamma^+\gamma_5\Phi(k)\big]$ and 
$S_T^i\,h_1(x){=}\int \mathrm dk^-\mathrm d^2\boldsymbol k_T
\tr\big[i\sigma^{i+}\Phi(k)\big]$. 
The operator $\mathcal U(0,\xi)$ is the gauge link,
a path-ordered exponential.
The gauge link is necessary to obtain a colour gauge-invariant definition of the quark-correlator.
Colour gauge-invariance of the quark-correlator is a prerequisite for the distribution functions that are defined via the quark-correlator to be gauge-invariant and have any physical relevance.

As stated above, the gauge link
$\mathcal U(0,\xi)\equiv\mathcal P\exp ig\int_C\mathrm dz\cdot A(z)$
is a path-ordered exponential connecting the fields at $0$ and $\xi$ along a certain path $C$.
This has important implications for processes where the intrinsic transverse momenta of the partons play a role.
If that is the case,
\ie, if one is sensitive to transverse directions,
the points $0$ and $\xi$ in the definition of the quark-correlator will also be separated in the transverse direction $\xi_T$.
If $\xi_T{=}0$ the gauge link runs along the light-cone,
if $\xi_T{\neq}0$ it must include a transverse piece (Brodsky \etal 2002, Belitsky \etal 2003),
leading to different possible integration paths.
The integration path $C$ that we have to take in a certain process is determined by the hard scattering part of that process.
In fact, it can explicitly be calculated by taking all interactions of the current quark with the spectator partons into account (Boer and Mulders 2000).
For example, 
in semi-inclusive deep inelastic scattering 
({\sc{sidis}}: $\ell+h_1\to\ell^\prime+h_2+X$)
the points $0$ and $\xi$ are connected to each other via light-cone future-infinity.
In Drell-Yan scattering
({\sc{dy}}: $h_1+\overline h_2\to\ell+\overline\ell+X$), 
however,
they are connected via light-cone past-infinity.
Due to the different gauge links in {\sc{sidis}} and {\sc{dy}},
the difference of the average transverse momentum of the quarks in the incoming proton in these two processes does not necessarily have to vanish.
Instead, it is proportional to an object that is called a \emph{gluonic pole}
(Boer \etal 2003).
The gluonic poles play a role in single spin asymmetries measured in experiment.

Recent studies have shown that scattering processes with more complicated partonic subprocesses than {\sc{sidis}} and {\sc{dy}} in general also have more complicated link structures (Bomhof \etal 2004).
One new effect is an extra winding of the integration path $C$ around light-cone future and past-infinity before connecting the points $0$ and $\xi$.
Another is the occurrence of traced Wilson loops in the quark-correlator.
All these new structures occur in hadronic pion production processes,
\eg $p^\uparrow p\to\pi X$ and $\overline p^\uparrow p\rightarrow\pi X$.
Such processes have been studied \eg at {\sc{fermilab}} and {\sc{rhic}}.
The exact form of the gauge links in these processes is now under investigation.
The newly found link structures will lead to enhancements of the gluonic poles,
but there is still work to be done in finding out the specific observables in which one sees their effects.
Other topics for future research are the implications that the new link structures have on issues such as universality and factorisation (Collins and Metz 2004).

\vspace{2mm}

I would like to thank the organizers of {\sc{sussp58}} and {\sc{i3hp}} for the successful summer school and workshop and for offering me the opportunity to present this talk. 
I would also like to thank the other participants of the summer school and workshop for helpful discussions.
Particularly Piet Mulders and Fetze Pijlman are acknowledged for their indispensable help.
This work was partially supported by the foundation for Fundamental Research of Matter ({\sc{fom}}) and the National Organization for Scientific Research ({\sc{nwo}}).

\vspace{-2mm}

\section*{References}
\vspace{-3mm}
\frenchspacing
\begin{small}
Feynman R P, 1969, \textit{Phys~Rev~Lett} \textbf{l23} 1415.\\
Ryckbosch D, 2004, Lectures presented at the {\sc{sussp58}} Summer School on Hadronic Physics\\ \hspace*{1cm}(to be published in the proceedings).\\
Soper D E, 1977, \textit{Phys Rev} \textbf{D15} 1141.\\
Brodsky S J, Hwang D S, Schmidt I, 2002, \textit{Phys Lett} \textbf{B530} 99.\\
Belitsky A V, Ji X, Yuan F, 2003, \textit{Nucl Phys} \textbf{B656} 165.\\
Boer D, Mulders P J, 2000, \textit{Nucl Phys} \textbf{B569} 505.\\
Boer D, Mulders P J, Pijlman F, 2003, \textit{Nucl Phys} \textbf{B667} 201.\\
Bomhof C J, Mulders P J, Pijlman F, 2004, \textit{Phys Lett} \textbf{B596} 277.\\
Collins J C, Metz A, 2004, hep-ph/0408249.
\end{small}
\nonfrenchspacing

\end{document}